\documentclass{article}
\usepackage{spconf,amsmath,graphicx}
\usepackage{amssymb}

\usepackage{multirow}
\usepackage{booktabs}
\usepackage{cite}

\title{TOLD: A Novel Two-stage Overlap-aware Framework for \\Speaker Diarization}
\name{Jiaming Wang$^*$, Zhihao Du$^*$, Shiliang Zhang\thanks{* Equal contribution.}}
\address{Speech Lab, Alibaba Group, China\\
\texttt{\{wangjiaming.wjm,neo.dzh,sly.zsl\}@alibaba-inc.com}}
\begin{document}
\ninept
\maketitle
\begin{abstract}
Recently, end-to-end neural diarization~(EEND) is introduced and achieves promising results in speaker-overlapped scenarios. In EEND, speaker diarization is formulated as a multi-label prediction problem, where speaker activities are estimated independently and their dependency are not well considered.
To overcome these disadvantages, we employ the power set encoding to reformulate speaker diarization as a single-label classification problem and propose the overlap-aware EEND~(EEND-OLA) model, in which speaker overlaps and dependency can be modeled explicitly. Inspired by the success of two-stage hybrid systems, 
we further propose a novel \textbf{T}wo-stage \textbf{O}ver\textbf{L}ap-aware \textbf{D}iarization framework~(TOLD) by involving a speaker overlap-aware post-processing (SOAP) model to iteratively refine the diarization results of EEND-OLA.
% , which considers dependency among different speakers and handles overlapping speech explicitly. 
% Specifically, in the first stage, we involve 
% In the second stage, a speaker overlap-aware post-processing model is employed to further refine the diarization results iteratively.
Experimental results show that, compared with the original EEND, the proposed EEND-OLA achieves a 14.39\% relative improvement in terms of diarization error rates~(DER), and utilizing SOAP provides another 19.33\% relative improvement. As a result, our method TOLD achieves a DER of 10.14\% on the CALLHOME dataset, which is a new state-of-the-art result on this benchmark to the best of our knowledge.
\end{abstract}
\begin{keywords}
speaker diarization, end-to-end neural diarization, two-stage framework, overlap-aware modeling
\end{keywords}
\section{Introduction}
\label{sec:intro}
Speaker diarization aims to solve the problem namely ``who spoken when'', which plays an essential role in many real-world applications, such as telephone transcription~\cite{telephone_transcription}, meeting minutes~\cite{meeting_minutes} and subtitle annotation~\cite{subtitle_annotation}. In these applications, accurate diarization results are important to achieve better system performance~\cite{multi_speaker_asr2}.

% In general, recent speaker diarization approaches can be divided into two groups: clustering-based methods and neural network based models. 
Conventional speaker diarization methods are based on clustering algorithms, in which voice activity detection~(VAD)~\cite{SAD} is first adopted to split the raw audio into speech segments. Then, speaker embedding extraction models, such as i-vectors~\cite{ivectors_1}, d-vectors~\cite{dvectors}, x-vectors~\cite{xvectors} and c-vectors~\cite{cvectors} are adopted to generate speaker embeddings for each segment. These embeddings are clustered to group segments belonging to the same speaker in an unsupervised fashion, such as K-means~\cite{kmeans_1}, spectral clustering~\cite{spectral_cluster_2} and agglomerative hierarchical clustering~(AHC)~\cite{ahc_1}. 
Due to the unsupervised manner, these methods do not minimize diarization errors directly resulting in the sub-optimal results. To solve the problem, supervised clustering methods are introduced \cite{LiK0W21,ZhangSWYY22}.
% One main disadvantage of clustering-based methods is that they are not optimized to minimize diarization errors directly as the clustering is processed in an unsupervised fashion. 
However, both supervised and unsupervised clustering methods can hardly deal with overlapping speech due to the speaker-homogeneous assumption~\cite{overlap_speech_1}. 
% The reason is that these methods usually assume that there is at most one speaker active at each time frame. 

% Another branch of speaker diarization researches is neural-based methods, one of which is EEND-based methods and we will focus on EEND-based methods in this paper. 
To deal with overlapping speech, end-to-end neural diarization~(EEND)~\cite{EEND} is introduced by formulating speaker diarization as a multi-label prediction problem. EEND models can minimize diarization errors directly with the permutation-invariant training~(PIT) loss function~\cite{PIT}. In SA-EEND~\cite{SA_EEND} and CB-EEND~\cite{CB_EEND}, multi-head self-attention~\cite{self_attention_1} blocks are employed to further improve the diarization performance.
% instead of original bidirectional long short-term memory ~(LSTM)~\cite{bi_lstm} blocks as the speech encoder. 
% However, both EEND and SA-EEND can only deal with a fixed number of speakers, which limits their applications. 
To handle a flexible number of speakers, encoder-decoder based attractor (EDA) \cite{EEND_EDA} is involved into EEND, which captures the information of the whole speech sequence and generates attractors for activated speakers. EEND-EDA can achieve better performance than EEND and SA-EEND even for a fixed number of speakers.

Recently, two-stage hybrid systems are introduced to utilize the advantages of clustering methods and EEND models. In \cite{tsvad, EEND-post, EAND}, clustering methods are employed as the first stage to obtain a flexible number of speakers, and then the clustering results are refined with neural diarization models as post-processing, such as two-speaker EEND, target speaker voice activity detection (TSVAD) and speaker overlap-aware neural diarization. Meanwhile, in \cite{EEND-global-local} and \cite{EEND-vector-clust}, the EEND-EDA model is employed as the first stage to extract attractors for each segment, and then clustering methods are used to obtain the inter-segment speaker correspondence by grouping the extracted attractors of each segment.

In this paper, we first propose the overlap-aware EEND~(EEND-OLA) model, in which the speaker dependency and overlaps are modeled explicitly by reformulating speaker diarization as a single-label classification problem with the power set encoding~(PSE).
% However, EEND-based methods usually formulate speaker diarization as a multi-label prediction problem, in which activities of each speaker are predicted independently. Such formulation can impair the diarization performance, especially for overlapping speech.
Then, we further propose a novel \textbf{T}wo-stage \textbf{O}ver\textbf{L}ap-aware \textbf{D}iarization framework~(TOLD), where a speaker overlap-aware post-processing (SOAP) model is involved to iteratively refine the results of overlap-aware EEND.
Specifically, in the first stage, an LSTM based EDA module is employed to extract attractors, and the corresponding order between speakers and attractors are determined by minimizing the permutation-invariant training loss function. Given order-determined attractors, we can represent speaker overlaps with a single label rather than a set of multiple binary labels. %  with the power set encoding (PSE).
% we introduce the power set encoding (PSE) to reformulate speaker diarization as a single-label classification problem, in which speaker overlaps are represented by a single label rather than a set of multiple binary labels.
% Specifically, 
In the second stage, we select non-overlapped speech segments to extract the initial speaker profiles according to the diarization results from the first stage. Then, speaker profiles and acoustic features are fed into the SOAP model to further refine the results by iteratively performing profile extraction and overlap-aware diarization.

\section{Two-stage Overlap-aware Diarization}
\begin{figure*}[htb]
  \centering
  \includegraphics[width=0.8\linewidth]{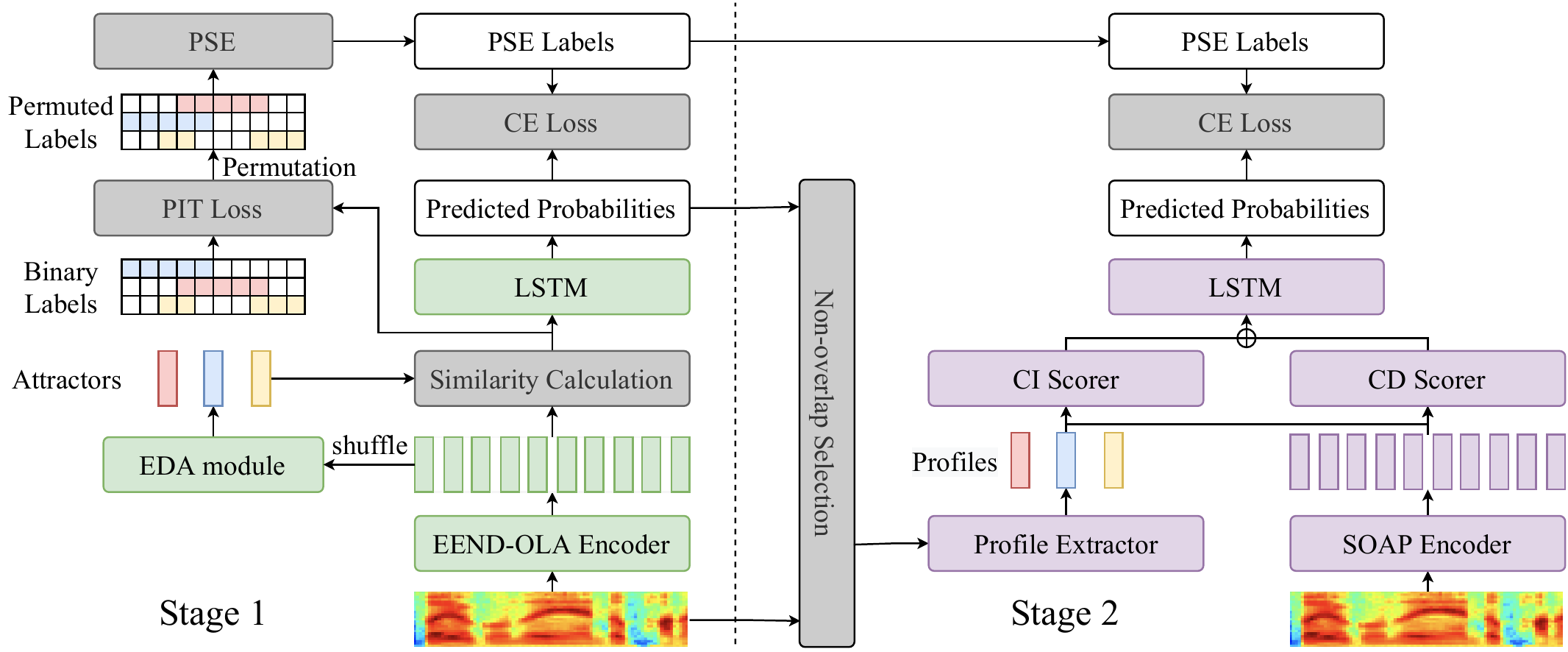}
  \caption{The overall architecture of the proposed method, TOLD, where $\oplus$ denotes concatenation operation.}
  \label{fig:overall}
\end{figure*}

In this section, we describe the proposed two-stage framework, TOLD. As shown in Fig.~\ref{fig:overall}, our system consists of two models named EEND-OLA and SOAP, respectively.

\subsection{Overlap-aware EEND}
\label{sec:first_stage}
At the first stage, we propose overlap-aware EEND~(EEND-OLA), which simplify the speaker diarization into a single-label classification problem with power set encoding~(PSE). As shown in Fig.~\ref{fig:overall}, the input is a $T$-length sequence of $F$ dimensional features $\mathbf{X}=[\mathbf{x}_{1},\ldots,\mathbf{x}_{T} ] \in \mathbb{R}^{F \times T}$. A transformer encoder without position encoding is utilized as a speech encoder to extract speech embeddings, which can be represented as $\mathbf{E}=[\mathbf{e}_{1},\ldots,\mathbf{e}_{T}] \in \mathbb{R}^{D \times T}$. Then, these embeddings are fed into an LSTM based EDA module to compute a flexible number of attractors $\mathbf{A}=[\mathbf{a}_{1},\ldots,\mathbf{a}_{S}] \in \mathbb{R}^{D \times S}$:
\begin{equation}
\begin{split}
    \mathbf{h}_{0}, \mathbf{c}_{0} &=\text{LSTMEncoder} \left( \mathbf{e}_{1}, \ldots,\mathbf{e}_{T} \right) \\
    \mathbf{a}_{s}, \left( \mathbf{h}_{s}, \mathbf{c}_{s} \right) &=\text{LSTMDecoder} \left(\mathbf{0}, \left(\mathbf{h}_{s-1}, \mathbf{c}_{s-1} \right) \right)
\end{split}
\end{equation}
where $1 \leq s \leq S+1$ and $S$ is the number of activated speakers in the sequence. $\mathbf{0}$ is a $D$-dimensional zero vector. $\mathbf{h}_{s-1}$ and $\mathbf{c}_{s-1}$ represent the hidden and cell states of LSTM from the previous step, respectively. As described in~\cite{EEND_EDA}, we randomly shuffle the speech embeddings across the time axis to improve the performance.

The training target of EEND-OLA is minimizing the weighted summation of the attractor existence loss, PIT loss and PSE loss. The attractor existence loss is applied to guide the model to generate the accurate number of attractors, which can be formulated as follows:
\begin{equation}
\begin{gathered}
    p_{s} =\frac{1}{1+\text{exp} \left( -\left( \mathbf{W}^{T}\mathbf{a}_{s} + \mathbf{b} \right) \right)}\\
    l_{s}=\begin{cases}
            1\quad s \in \left\{1, \ldots, S \right\} \\
            0 \quad s=S+1
          \end{cases}\\
    \mathcal{L}_{a}=\frac{1}{S+1} \text{BCE} \left(\mathbf{l}, \mathbf{p} \right)
\end{gathered}
\end{equation}
where $\mathbf{l}=\left[ l_{1},\ldots,l_{S+1} \right] ^ {T}$, $\mathbf{p}=\left[ p_{1},\ldots,p_{S+1} \right] ^ {T}$ and BCE denotes the binary cross entropy. $\mathbf{W}$ and $\mathbf{b}$ are trainable parameters.

The PIT loss is used to determine the corresponding order between attractors and speakers, which can be represented as follows:
\begin{equation}
\begin{gathered}
    \mathcal{L}_{d}=\frac{1}{T} \mathop {\min }\limits_{\phi \in \text{perm} \left(1,\ldots,S \right)} \sum \limits _{t=1}^{T} \text{BCE}\left(\mathbf{y}_{t}^{\phi},\hat{\mathbf{y}}_{t} \right) \\
    \hat{\mathbf{y}}_{t} = \sigma \left ( \mathbf{A}^{T} e_{t} \right) \in {\left( 0, 1 \right)}^{S} \\
    \mathbf{y}_{t} ={ \left[y_{t,1},\ldots,y_{t,S} \right] } \in {\left( 0, 1 \right)}^{S}
\end{gathered}
\end{equation}
where $\hat{\mathbf{y}}_{t}$ denotes the posterior probabilities and $\mathbf{y}_{t}$ denotes the ground-truth binary labels. By minimizing the PIT loss, the speaker order is determined and the corresponding labels are represented as $\mathbf{\widetilde{y}}_{t} = \mathbf{y}^{\phi^*}_{t} ={ \left[\widetilde{y}_{t,1},\ldots,\widetilde{y}_{t,S} \right] } \in {\left( 0, 1 \right)}^{S}$.

As for PSE loss, we calculate the PSE labels from the order-determined binary labels $\mathbf{\widetilde{y}}_{t}$ as follows:
\begin{equation}
    w_{t}=\text{PSE}(\mathbf{\widetilde{y}}_{t})=\widetilde{y}_{t,1} \times 2^{0} +,\ldots,+\widetilde{y}_{t,S} \times 2 ^ {S-1} 
\end{equation}
According to real scenarios, we assume that there are at most $K=3$ speakers active simultaneously in one frame and the total number of categories can be computed as:
\begin{equation}
    N=\sum_{k=0}^{K} \binom{S_{\text{max}}}{k} =\sum_{k=0}^{K} \frac{S_{\text{max}}!}{k!(S_{\text{max}}-k)!}
\end{equation}
where $S_{\text{max}}$ represents the predefined maximum number of speakers.
In this way, speaker diarization is reformulated as a single-label classification problem. To predict PSE labels, we introduce an additional LSTM layer to make full use of contextual information. Since the number of attractors is flexible and the LSTM layer requires the input to be a sequence of fixed-dimensional vectors, we pad attractors as the number of $S_{\text{max}}=8$ with zero vectors. The similarities between padded attractors and speech embeddings are obtained by inner products and fed into the LSTM layer to predict the probabilities of PSE labels $\hat{\mathbf{w}}_t \in \mathbb{R}^{N}$: 
% \begin{equation}
% \begin{split}
%     \mathbf{d}_t = [<\mathbf{e}_t,\mathbf{a}_1&>,<\mathbf{e}_t,\mathbf{a}_2>,\dots,<\mathbf{e}_t,\mathbf{a}_{S_\text{max}}>] \\
%     \hat{\mathbf{w}}_t &= \text{LSTM}(\mathbf{d}_{t};(\mathbf{h}_{t-1},\mathbf{c}_{t-1}))
% \end{split}
% \end{equation}
\begin{equation}
\begin{gathered}
    \mathbf{d}_t = [<\mathbf{e}_t,\mathbf{a}_1>,<\mathbf{e}_t,\mathbf{a}_2>,\dots,<\mathbf{e}_t,\mathbf{a}_{S_\text{max}}>] \\
    \hat{\mathbf{w}}_t =\text{Softmax}( \mathbf{W}^{T}(\text{LSTM}(\mathbf{d}_{t};(\mathbf{h}_{t-1},\mathbf{c}_{t-1})))+\mathbf{b})
\end{gathered}
\end{equation}
% where $\mathbf{h}_{t-1}$ and $\mathbf{c}_{t-1}$ represent the hidden and cell states of LSTM from the previous step, respectively. 
% $\mathbf{C}={\left[\mathbf{c}_{1},\ldots,\mathbf{c}_{T} \right]}^{T} \in \mathbb{R}^{T \times S_{\text{max}}}$. 
% Instead of predicting whether each speaker is active at each frame, PSE predicts the diarization results based on the hidden states of LSTM through a fully connected layer. 
where $\hat{\mathbf{w}}_t$ represents the prediction of PSE labels. Then, the PSE loss can be obtained as follows:
\begin{equation}
    \mathcal{L}_{\text{PSE}} = \frac{1}{T}\sum_{t=1}^{T} \text{CE}\left(\hat{\mathbf{w}}_t, w_{t}\right)
\end{equation}
where CE denotes the cross entropy loss.
% where $w_{t}$ represents the ground truth PSE label and $\mathbf{h}_{t}$ is the output of LSTM at the frame $t$.
The total loss at the first stage can be represented as follows:
\begin{equation}
    \mathcal{L}_{\text{Stage1}}=\mathcal{L}_{\text{PSE}} + \mathcal{L}_{d} + \alpha \mathcal{L}_{a}
    \label{eq:pse_loss}
\end{equation}
In this paper, $\alpha$ is set to 1.0 and 0.01 for pre-training and data adaption, respectively.

\subsection{Speaker overlap-aware post-processing}
In SOAP, non-overlapped speech segments are selected and fed into a pre-trained x-vector extractor to obtain the speaker profiles $\mathbf{V}=[\mathbf{v}_1,\mathbf{v}_2,\dots,\mathbf{v}_S]$. The ResNet34 \cite{HeZRS16} is employed as our embedding extraction model, which is optimized by minimizing the ArcFace loss function \cite{DengGXZ19} with a margin of 0.25 and softmax pre-scaling of 8. The encoding layer is based on global statistic pooling, and the dimension of the speaker embedding layer is 256. More details can be found in \cite{EAND}. Meanwhile, the speaker embedding extractor is also used to initialize the speech encoder, which consists of convolutional blocks (Conv), windowed statistic pooling (SP) and embedding layer (Emb).
Note that, different from embedding extractor, the statistic pooling of speech encoder is calculated on a window rather then the entire input sequence:
\begin{equation}
		\mathbf{h}_t =\text{Emb}(\text{SP}(\text{Conv}(\mathbf{X}) \mathbf{I}_{t-2/l:t+l/2}))
\end{equation}
where $\mathbf{I}_{t-2/l:t+l/2}$ represents an identity window with ones from $t - l/2$ to $t + l/2$ and zeros otherwise. $\mathbf{h}_t$ denotes the outputs of embedding layer in speech encoder at time-step $t$.
% The window length $l$ is set to 20 in our experiments.

Given encoded features $\mathbf{H}=[\mathbf{h}_1,\mathbf{h}_2,\dots,\mathbf{h}_T]$ and speaker profiles $\mathbf{V}$, context-dependent (CD) and context-independent (CI) scorers are employed to predict activities of speaker $s$ at time-step $t$:
\begin{equation}
\begin{split}
	\mathbf{S}^{CI}_{s,t}&=\text{DNN}\left(\mathbf{v}_s\oplus\mathbf{h}_t\right)
\end{split}
\end{equation}
where DNN represents a deep neural network consisting of three fully-connected layers with Tanh activation function and an output layer with 256 units. $\oplus$ means the concatenation of two vectors.
In CD scorer, a multi-head self attention (MHSA) based neural network is employed to predict the context-dependent probabilities of speaker $s$ at all time-steps:
\begin{equation}
\begin{split}
	&z_{s,0} = [\mathbf{h}_1\oplus\mathbf{{v}}_s, \mathbf{h}_2\oplus\mathbf{{v}}_s,\dots,\mathbf{h}_T\oplus\mathbf{{v}}_s] \\
	&\bar{z}_{s,l} = z_{s,l-1} + \text{MHSA}_l(z_{s,l-1}, z_{s,l-1}, z_{s,l-1}) \\
	&z_{s,l} = \bar{z}_{s,l} + \text{max}(0, \bar{z}_{s,l}\mathbf{W}^l_1+\mathbf{b}^l_1)\mathbf{W}^l_2+\mathbf{b}^l_2 \\
	&\mathbf{S}^{CD}_{s} = \text{Sigmoid}(\mathbf{W}^o z_{s,L^{CD}} + \mathbf{b}^{o}) 
\end{split} 
\end{equation}
where $\text{MHSA}_l(Q,K,V)$ represents the multi-head self attention of the $l$-th layer \cite{self_attention_1} with query $Q$, key $K$, and value $V$ matrices.
$\mathbf{W}^l_*$ and $\mathbf{b}^l_*$ denotes the learnable weight and bias of the $l$-th layer ($o$ for output layer). 

We employ a LSTM layer to model speaker dependency and overlaps.
The inputs of LSTM are concatenated scores from CI and CD scorers, and the target outputs are PSE labels given the speaker order in profiles:
% \begin{equation}
%     \hat{\mathbf{w}}_t=\text{LSTM}\left(\mathbf{S}^{CI}\oplus\mathbf{S}^{CD}\right)
% \end{equation}
\begin{equation}
    \hat{\mathbf{w}}_t=\text{Softmax}(\mathbf{W}^{T} ( \text{LSTM}(\mathbf{S}^{CI}\oplus\mathbf{S}^{CD})) + \mathbf{b})
\end{equation}
% where $\hat{\mathbf{w}}_t$ represents the prediction of PSE labels.

We adopt a multi-task learning strategy to optimize our model.
The main training objective is minimizing the CE loss between predicted probabilities of PSE labels $\hat{\mathbf{w}}_t$ and their ground-truth counterparts $w_t$:
\begin{equation}
	\mathcal{L}_{\text{CE}} = \frac{1}{T}\sum_{t=1}^{T}{\text{CE}\left(\hat{\mathbf{w}}_t, w_t\right)}
\end{equation}
In addition, we provide extra guidance to the intermediate CI and CD scorers by minimizing the BCE loss between frame-level scores $\mathbf{S}^{*}_{t}$ and posterior probabilities of speaker activities $\mathbf{y}_{t}$:
\begin{equation}
    \mathcal{L}_{\text{Guide}}=\frac{1}{T}\sum_{t=1}^{T}{\text{BCE}\left(\mathbf{S}^{CI}_{t},\mathbf{y}_{t}\right)+\text{BCE}\left(\mathbf{S}^{CD}_{t},\mathbf{y}_{t}\right)}
\end{equation}
Finally, the training object of the second stage is obtained as:
\begin{equation}
    \mathcal{L}_{\text{Stage2}}=\mathcal{L}_{\text{CE}} + \lambda\mathcal{L}_{\text{Guide}}
\end{equation}
where $\lambda$ is a hype-parameter to balance the CE and guidance losses. According to preliminary experiments, $\lambda$ is set to $0.1$ in this paper.

\section{Experimental Settings}
% To make a fair comparison, we follow the EEND-EDA framework and also shuffle speech embedding order to help EDA generate better attractors in the first stage, which helps improve performance as described in the EEND-EDA. However, the shuffle strategy makes position encoding unsuitable in transformer encoder, leading to the model lack of context information. On the contrary, we remove EDA at the second stage and directly use the attractors generated in the first stage. Therefore, we apply the chronological order, which makes utilizing context information
% As we employ transformer encoder without position encoding, the extracted speech embeddings lack of context information. The reason is that we follow the EEND-EDA framework, which needs to shuffle speech embeddings to help EDA learn better and improve performance.
\subsection{Data}
We first pre-train the models with simulated mixtures and then finetune them on the real dataset.
The simulated mixtures are generated from Switchboard-2 (Phase \uppercase\expandafter{\romannumeral1} \& \uppercase\expandafter{\romannumeral2} \& \uppercase\expandafter{\romannumeral3}), Switchboard Cellular (Part 1 \& 2) and the NIST Speaker Recognition Evaluation (2004 \& 2005 \& 2006 \& 2008) datasets. Noises from MUSAN~\cite{musan} and simulated room impulse responses~\cite{rirs} are applied to make simulated mixtures more similar to real conversations. To train the models at the first stage, we adopt the simulation procedure described in~\cite{SA_EEND} and obtain 400,000 mixtures with the speaker number varied from one to four.
For the second stage, we follow the simulation procedure in \cite{EAND} and obtain 112,500 simulated mixtures for training with the speaker number varied from two to seven.

To evaluate our method in the real-world scenarios, we employ the widely used CALLHOME dataset, which comprises 500 recordings with the speaker number varied from two to seven speakers in each recording. For fair comparison, we adopt the same evaluation policies as described in \cite{EEND_EDA}. The entire dataset is split into two subsets for adaption and evaluation, and each subset includes 250 recordings.

\subsection{Model settings and training details}
\label{subsec:setting}
The input of TOLD is log-scaled Mel-filterbank features with the window size of 25ms and the shift of 10ms. The EEND-OLA and SOAP encoders consist of four transformer blocks, where each block comprises a multi-head attention layer with 256 attention units and four heads. As for the EDA module, two unidirectional LSTM layers are employed as attractor encoder and decoder, respectively. To model the contextual information, we employ a LSTM layer to predict the posterior probabilities of PSE labels at both stages. All LSTM layers consist of 256 hidden units and cell states.

% We evaluate the proposed methods on both simulated mixtures and CALLHOME. For simulated mixtures, following \cite{EEND_EDA}, 
At the first stage, we train EEND-OLA for 100 epochs on 2-speaker simulated mixtures with the maximum sequence length of 50 seconds and finetune the model for 25 epochs on all simulated mixtures with speaker number from one to four. Since recordings of test set usually longer than 50 seconds leading to a mismatch between training and test, we further finetune the model for another epoch on all mixtures with the maximum sequence length of 200 seconds.
At the second stage, the original single-speaker utterances from Switchboard-2, Switchboard Cellular and NIST Speaker Recognition Evaluation datasets are used to train the profile extractor, which is further finetuned on the real data with non-overlap segments.
We also train the SOAP model for 100 epochs, in which the SOAP encoder is initialized with the pre-trained profile extractor and frozen in the first 50 epochs. The sequence length is limited to 16 seconds.
To evaluate on CALLHOME dataset, we finetune the EEND-OLA model for 100 epochs on the adaption set with the maximum sequence length of 200 seconds and average the parameters in the last ten checkpoints. For SOAP, we finetune the model for ten epochs and the last five checkpoints are averaged. We employ the diarization error rate (DER) as our evaluation metric with a collar tolerance of 0.25 seconds.

\section{Experimental results}
% In this section, we first evaluate the effect of overlap-aware modeling for EEND-OLA by comparing with other one-stage models. Then, EEND-OLA is involved into the two-stage framework, TOLD, and compared with other two-stage systems. Furthermore, an ablation study is performed for TOLD to evaluate the impact of different models at two stages.
\subsection{The effect of overlap-aware modeling}
We compare the proposed EEND-OLA with other one-stage methods on real-life conversations from the CALLHOME dataset, which includes a flexible number of speakers. The results are shown in Table~\ref{tab:single_callhome_der}. Note that the results of VBx are obtained with the oracle voice activity dection, while other methods use the model outputs directly. EEND-EDA(Paper) represents the results reported in~\cite{EEND_EDA}, and EEND-EDA denotes our own implementation with the training strategy mentioned in Section~\ref{subsec:setting}. Since, our implemented EEND-EDA achieves a better performance than the original version, we use it in the following experiments.

From Table~\ref{tab:single_callhome_der}, we can see that the proposed EEND-OLA achieves the best performance for a flexible number of speakers than other one-stage models.
Although SA-EEND and CB-EEND are designed for a fixed number of speakers, our method still achieves a comparable performance with them on the two-speaker recordings.
Compared with EEND-EDA, the proposed EEND-OLA achieves a 13.49\% relative reduction on the averaged DER, which reveals the effectiveness of overlap-aware modeling.
In addition, we also provide the diarization results derived from the intermediate states of EEND-OLA before PSE. As seen in the row ``--Before PSE'', the intermediate results are also better than EEND-EDA. This indicates that the PSE labels can improve the learning process of EEND-EDA.

\begin{table}[t!]
  \caption{Comparison of different one-stage systems in terms of DER(\%) on CALLHOME dataset. $\dagger$ means the oracle VAD is used.}
  \vspace{0.1cm}
  \label{tab:single_callhome_der}
  \centering
  \setlength\tabcolsep{2.5pt}
  \begin{tabular}{l c c c c c c}
    \toprule 
    \multirow{2}{*}{Method} & \multicolumn{6}{c}{Number of Speakers}  \\
    \cmidrule(lr){2-7}
    & 2  & 3 & 4 & 5 & 6 & All \\
    \midrule
    VBx$^\dagger$\cite{vbx}                   & 10.00   & 14.28   & 21.19 & 27.91 & 35.59 & 15.80    \\
    % SA-EEND~\cite{SA_EEND}                         & 8.56    & -   & -	& -	& -	& -    \\
    % CB-EEND~\cite{CB_EEND}                         & \textbf{6.82}    & -   & -	& -	& -	& -    \\
    EEND-EDA~\cite{EEND_EDA}        & 8.50    & 13.24   & 21.46 & 33.16 & 40.29 & 15.29    \\
    % EEND-EDA                & 7.79    & 13.54   & 19.65	& 30.04	& 35.12	& 14.53    \\
    \midrule
    EEND-OLA     & 6.91	    & \textbf{11.19}	& \textbf{17.08}	& \textbf{27.82}	& \textbf{30.95}	& \textbf{12.57}    \\
    % ~~--Before PSE    & 6.99	    & 12.22	& 18.84	& 29.18	& 35.27	& 13.59    \\
    \midrule
    \midrule
    % EEND-post~\cite{EEND-post}                       & 9.87    & 13.11	& 16.52	& 28.65	& 27.83	& 14.06    \\
    EEND-vector clust.~\cite{EEND-vector-clust}                 & 7.94    & 11.93	& 16.38	& \textbf{21.21}	& 23.10	& 12.49    \\
    EEND-global-local~\cite{EEND-global-local}               & 7.11    & 11.88	& 14.37	& 25.95	& 21.95	& 11.84    \\
    \midrule
    TOLD                  & \textbf{5.73}	    & \textbf{10.31}	& \textbf{11.96}	& 23.89	& \textbf{20.39}	& \textbf{10.14}    \\
    \bottomrule
  \end{tabular}
\end{table}

\subsection{Comparison of two-stage systems}
In Table~\ref{tab:hybrid_callhome_der}, we compare the proposed TOLD with other two-stage systems on real-life recordings from the CALLHOME dataset. In EEND-post, a 2-speaker end-to-end neural diarization model is employed as post-processing to refine the clustering results. Meanwhile, in EEND-vector clustering and EEND-global-local, long-term speaker-overlapped recordings are split into segments and processed by the end-to-end neural diarization model independently. Then, a clustering algorithm is used to obtain the speaker correspondence of each segment and aggregate the diarization results.
As we can see, the above two-stage methods are actually implemented by combining EEND models and clustering algorithms. Different from them, both of the tow stages in TOLD are based on neural network models, which provide the potential for unified end-to-end system.
In total, our TOLD achieves a 10.14\% DER on the CALLHOME test set, which is a new state-of-the-art result on this commonly-used benchmark.
% best performance in terms of averaged DER
% TOLD denotes the results of our proposed method which integrates two end-to-end methods. It is obvious that our proposed method can achieve significant improvement than other two-stage methods.
% are  propose to utilize advantages of end-to-end models and clustering-based models to deal with long and overlapped recordings. EEND-global-local introduce an unsupervised clustering process to help the end-to-end model handle the case of the larger speaker number than training. 

\begin{table}[t!]
  \caption{Comparison of different two-stage systems in terms of DER(\%) on CALLHOME dataset.}
  \vspace{0.1cm}
  \label{tab:hybrid_callhome_der}
  \centering
  \setlength\tabcolsep{2.5pt}
  \begin{tabular}{l c c c c c c}
    \toprule 
    \multirow{2}{*}{Method} & \multicolumn{6}{c}{Number of Speakers}  \\
    \cmidrule(lr){2-7}
    & 2  & 3 & 4 & 5 & 6 & All \\
    \midrule
    EEND-post~\cite{EEND-post}                       & 9.87    & 13.11	& 16.52	& 28.65	& 27.83	& 14.06    \\
    EEND-vector clust.~\cite{EEND-vector-clust}                 & 7.94    & 11.93	& 16.38	& \textbf{21.21}	& 23.10	& 12.49    \\
    EEND-global-local~\cite{EEND-global-local}               & 7.11    & 11.88	& 14.37	& 25.95	& 21.95	& 11.84    \\
    \midrule
    TOLD                  & \textbf{5.73}	    & \textbf{10.31}	& \textbf{11.96}	& 23.89	& \textbf{20.39}	& \textbf{10.14}    \\
    \bottomrule
  \end{tabular}
  \vspace{-0.1cm}
\end{table}

\begin{table}[t!]
  \vspace{-0.2cm}
  \caption{Comparison of different methods for two stages on the CALLHOME dataset. $\dagger$ means the oracle VAD is used.}
  \vspace{0.1cm}
  \label{tab:callhome_der_compare}
  \centering
  \setlength\tabcolsep{2.5pt}
  \begin{tabular}{l c c c c c c c}
    \toprule 
    \multirow{2}{*}{Stage1} & \multirow{2}{*}{Stage2} & \multicolumn{6}{c}{Number of Speakers}  \\
    \cmidrule(lr){3-8}
    & & 2 & 3 & 4 & 5 & 6 & All \\
    \midrule
    \multirow{2}{*}{VBx$^\dagger$}  & TSVAD         & 6.84    & \textbf{9.85}    & 14.46   & 24.02 & 31.50 & 11.38    \\
      & SOAP          & 6.26    & 9.91    & 14.22.  & 23.96 & 31.18 & 11.13    \\
    \midrule
    \multirow{2}{*}{EEND-EDA}  & TSVAD    & 6.25	  & 10.79   & 15.10   & 23.65 & 23.73 & 11.28    \\
    % & SOAP     & \textbf{5.62}    & 10.87   & 12.82   & 25.59 & 20.82 & 10.58    \\
    & SOAP & \textbf{5.73} & 10.57 & 14.67 & 24.15 & 22.53 & 10.90 \\
    \midrule
    \multirow{2}{*}{EEND-OLA}~~ & TSVAD~~    & 6.23 & 10.43 & 12.49 & 23.96 & 21.67 & 10.54 \\
    & SOAP & \textbf{5.73} & 10.31 & \textbf{11.96}   & \textbf{23.89} & \textbf{20.39} & \textbf{10.14}    \\
    \bottomrule
  \end{tabular}
  \vspace{-0.3cm}
\end{table}

\subsection{Ablation study of TOLD}
To further investigate the impact of different methods for the two stages, we design ablation experiments, and the results are shown in Table~\ref{tab:callhome_der_compare}. For the first stage, we evaluate VBx, EEND-EDA and EEND-OLA. For the second stage, TSVAD and SOAP are compared. 
According to the results in Table~\ref{tab:single_callhome_der} and Table~\ref{tab:callhome_der_compare}, we find that, with the same second-stage model, the lower DER provided by the first stage usually means the better performance achieved by the whole system. No matter which second-stage model is used, the proposed EEND-OLA always provides a better performance than VBx and EEND-EDA.
Besides, no matter which model is used for the first stage, SOAP provides a lower DER than TSVAD. This may indicates that SOAP is more appropriate for post-processing.

\section{Conclusions}
In this paper, we reformulate speaker diarization from a multi-label predication problem into a single-label classification problem. Through this formulation, we propose the EEND-OLA model to predict the power set encoded labels, which is derived from the order-determined attractors and multiply binary labels. By explicitly modeling the speaker dependency and overlaps, not only the outputs but also the intermediate states of EEND-OLA can achieve a better diarization performance than the original EEND-EDA model.
% We find that such overlap-aware modeling improves not only the diarization performance but also the 
Inspired by the recent success of two-stage hybrid systems, we further propose the TOLD framework, in which the initial diarization results from EEND-OLA is iteratively refined by the overlap-aware post-processing, SOAP. 
By comparing the combinations of different methods for two stages, we find that the proposed EEND-OLA can provide a better initial diarization results for the second stage than VBx and EEND-EDA. In addition, SOAP is a better post-processing method than TSVAD, no matter which method is used at the first stage.
Finally, our TOLD achieves a new state-of-the-art results on the commonly-used CALLHOME dataset, which indicates the effectiveness of overlap-aware modeling via PSE.
% In this paper, we propose TOLD, a novel two-stage overlap-aware framework for speaker diarization, which can reduce the training difficulty and handle the overlap speech ambiguity problem. In the first stage, we propose mPIT to help generate more accurate attractors. In the second stage, the attractors generated in the first stage is employed and the permutation of speaker labels has been determined, which reduces the training difficulty. Besides, PSE is introduced to help obtain the diarization results based on a learnable module with the similarities between attractors and speech embeddings, instead of predicting the activity of each speaker separately, which can alleviate the overlap speech ambiguity problem. Our proposed method outperforms EEND-EDA and x-vector-based clustering methods on both simulated mixtures and CALLHOME datasets. 

\bibliographystyle{IEEEbib}
\bibliography{refs}

\end{document}